# GPU-Framework for Teamwork Action Recognition

Mohamed H. Elhoseiny[1], H.M. Faheem[2], T.M. Nazmy[2], Eman Shaaban[2]
m.elhoseiny@cs.rutgers,edu, hmfaheem@ieee.org, tmnazmy@cis.asu.edu.eg, eman.shaaban@cis.asu.edu.eg
[1] Rutgers University, Computer Science Department
[2] Faculty of Computer and Information Sciences, Ain Shams University, Abbassia, Cairo, Egypt.

**Abstract**

*Real time processing for teamwork action recognition is a challenge, due to complex computational models to achieve high system performance. Hence, this paper proposes a framework based on Graphical processing Units (GPUs) to achieve a significant speed–up in the performance of role based activity recognition of teamwork. The framework can be applied in various fields, especially athletic and military applications. Furthermore, the framework can be customized for many action recognition applications. The paper presents the stages of the framework where GPUs are the main tool for performance improvement. The speedup is achieved by performing video processing and Machine learning algorithms on GPU. Video processing and machine learning algorithms covers all computations involved in our framework. Video processing tasks on involves GPU implementation of Motion detection, segmentation and object tracking algorithms. In addition, our framework is integrated with GPUCV, a GPU version of OpenCV functions. Machine learning tasks are supported under our framework with GPU implementations of Support Vector Machine (SVM) for object classification and feature discretization, Hidden Marcov Model (HMM) for activity recognition phase, and ID3 algorithm for role recognition of team members. The system was tested against UC-Teamwork dataset and speedup of 20X has been achieved on NVidia 9500GT graphics card (32 500MHZ processors).*

## 1. Introduction

GPU has been one of the state-of-the-art technology used to solve performance problem due to its intensive computational power, beside its low cost that might make simple personal computer performs as powerful as super computer by attaching GPU chip. One of the leading companies that have developed this chip is NVidia Corporation.

NVidia presents an architecture known as CUDA provided with a SDK, accessible to software developers through standard programming languages. In the same direction, OpenCL, initially developed by Apple Incorporation, became a standard framework for writing Kernels on heterogeneous platforms that consists of CPUs and GPUs. As a result, advances in GPU industry and development have been achieved and a lot of work has been done to get GPU applied in many high performance fields. Examples include Bioinformatics, Image processing and computer vision.

Applications of image processing include Hyper-spectral image processing [1], where Javier Setoain et al presented an investigation of multi-level parallel implementations. In Computer vision field, Jean-Philippe et al presented GPUCV [2] as GPU implementation of popular OpenCV functions. This work has achieved improvement of the speedup in computer vision applications, however optimality is not guaranteed due to generality of these functions. As a rule, special purpose GPU algorithm design mostly results in a big step in performance. In addition, GPUCV handeles openCV functions in a general way, which motivates a lot of people to build special purpose implementations on GPU.

Moreover, Peihua Li et al presented parallel implementation of mean shift tracking [3] on GPU as six kernel functions. They used of K-means clustering to partition color space into distribution of small pins; consequently all key components were mapped onto GPU with great speedup.

In the same direction, a lot of research has been done towards enhancing performance of detection, tracking and segmentation phases. Furthermore, a noticeable effort has been exerted to achieve significant improvement in the implementation of machine learning algorithms, due to its great contribution in various computer vision applications. In this direction, Chan Lui presented GPU implementations of Time Series Models [12] and HMM [4]. Similarly, Austin Carpenter presented GPU implementation of SVM by Austin Carpenter [5].

Currently, GPUs are widely deployed as a powerful resource to achieve intensive computing tasks in high speed. Since Action Recognition systems requires huge computational power to fit in real-time.

In this paper, we are trying to build a linkage between GPUs and teamwork activity recognition systems as a framework, on which many real-time video-processing systems can be implemented. This framework targets the developer of action recognition systems. Through the rest of paper, we will refer to the target users of the framework as developers. Our framework is mainly software that integrates some of the existing approaches on GPU to perform action recognition. We aim to develop and maintain this framework in the future so that it could be used to address large-scale video processing applications including activity recognition. This paper is organized as follows. Section 2 presents the proposed framework. Section 3 shows a prescription to configure our framework to achieve high performance on a general hardware. Section 4 shows how the framework could be configured for any system. Part 5 presents experimental results that report the achieved speedup.

## 2. GPU TEAMWORK ACTIVITY RECOGNITION FRAMEWORK

For building a channel between GPUs and activity recognition systems, we investigated many major techniques used in action recognition and computer vision, with the objective of making this connection results in an initial framework, on which many action recognition systems can be built to gain speedup. Figure 1 presents the workflow of the proposed GPU framework. The input is raw images are captured of moving camera, and the output is the recognized activity. The colored modules are the ones with general GPU blocks to be configured according to application requirements. The non-colored modules, (Agent oriented feature extraction, Team oriented feature extraction), are implemented by the developer, according to the potential application running under the framework. However the non- colored modules are implemented by the developer, the framework provides the developer with GPUCV functions which gives him support to achieve speedup with minimal effort. Furthermore, the framework accepts the developer implementation either on GPU or CPU.

As illustrated in Figure 1, the framework spans all stages required for many action recognition systems with specifications of each component, defined on the arrows.

Besides, the framework has been built, such that functionality of GPUCV (i.e. GPU version of Intel OpenCV functions) [4] is supported and integrated as an infrastructure that can be used by the developer. Next sections present the detailed implementation of each component in the framework that together comprises the framework.

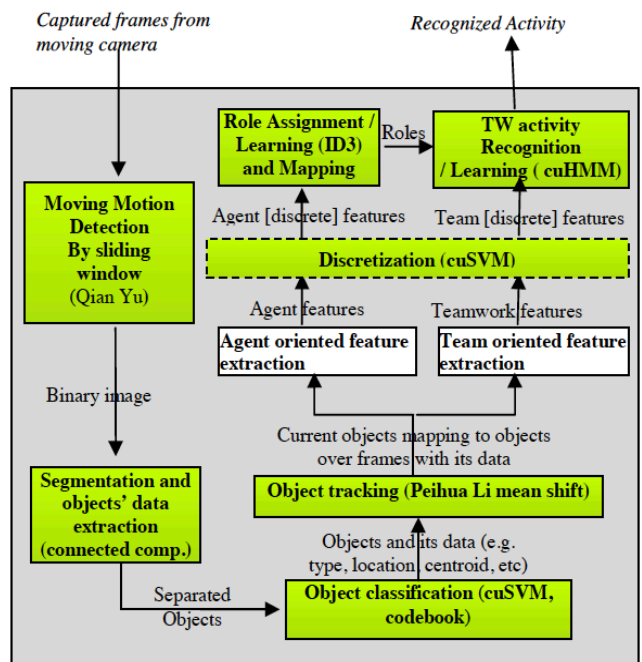

Figure 1: Process flow of our GPU framework

### 2.1. Moving Motion Detection by sliding window

Qian Yu et al [6] achieved significant acceleration of motion detection of moving framework based on GPU, which leads to building background model and detecting motion regions at 18fps 320x240 videos. The process of detection on GPU was implemented as two steps, warping images and computing the background model. The main objective was to minimize memory transferring between GPU and CPU.

Figure 2 illustrates how the structure stores frames of the window as 2D textures in GPU memory. Next step is wrapping, which is performed on each frame of the window using GPU. Finally, the output of wrapping is passed to GPU component for collecting statistics from the correspondences of the window, which leads to background model.

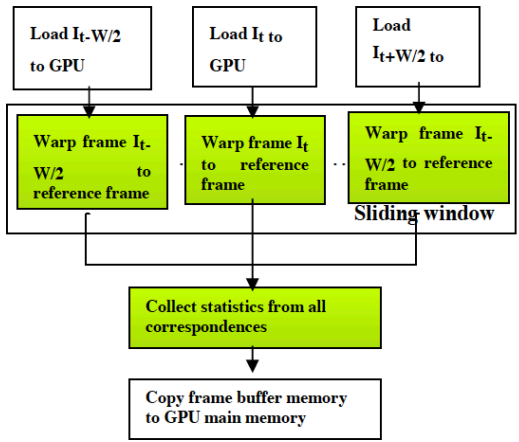

Figure 2: Quan Yu et al Motion detection on GPU (W is the window size of the frames required by the algorithm).

## 2.2. Segmentation and Objects' Data Extraction

In this stage, a GPU connected component algorithm is applied to segment objects in the scene (Figure 3). The speedup is gained by dividing image into N blocks then merging the computations is as detailed in [7]. But, our experiments show that the cost of this algorithm depends on the size and complexity of the labeled objects, so this technique is significantly not faster than the CPU classical 2-passes sequential algorithm achieved by using Intel OpenCV version but even slower.

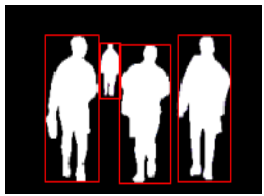

Figure 3: Segmentation and object extraction

## 2.3. Object classification (cuSVM, codebook)

This stage of processing is responsible for categorizing detected objects into classes. Speedup of this stage was achieved by a multilevel implementation. For instance, Jianpeng Zhou et al presented codebook algorithm to classify humans [8], While Vili Kellokumpu et al used SVM in [9] to classify human postures (Figure 4). So codebook can be used to ensure filtering out non- human objects, then human blobs is presented to SVM to classify human postures. Consequently, GPU implementations of these models are included as a basic part of our framework [5].

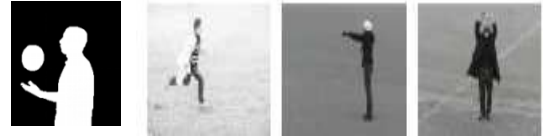

(a)          (b)

Figure 4: (a) Classification of object as a ball and human (b) Classification of different human postures.

## 2.4. Object tracking

In this component correspondence mapping of blobs across frames are constructed. We integrated the GPU implementation of Peihua Li et al's work [3], which was based on the idea of partitioning the color space of objects over the GPUs. This leads to a quite small number of histogram bins representing color distribution. That histogram was the base component of mean shift tracking algorithm. The speed-up of object tracking in the framework is achieved by this GPU implementation of mean-shift algorithm. Figure 5 shows an instance run of the tracking algorithm that can handle tracking many objects simultaneously due to the speed of the GPU implementation.

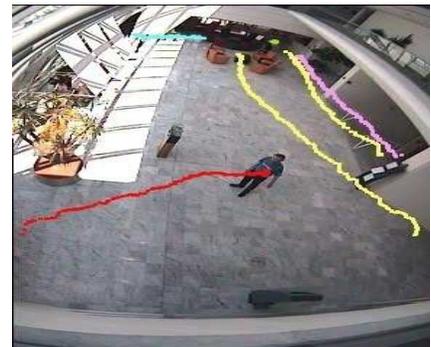

Figure 5: Object Tracking

## 2.5. Discretization(cuSVM)

After the detection and classification of objects, the next step is to handle features into HMM for recognition of actions in discrete form which to be plugged into HMMs. because discrete HMM is more accurate than continuous HMM as known in the literature [?], This phase encapsulate discretization of agent features (e.g. postures into a set of predefined ones, velocity as fast or slow, ... etc.), and team oriented features (e.g. cohesion as separated or merged) as illustrated in Figure 6.

4323

In this phase, the framework provides the developer with GPU implementation of SVM to discretize features in . Moreover, the developer can integrate his own discretizer into the framework.

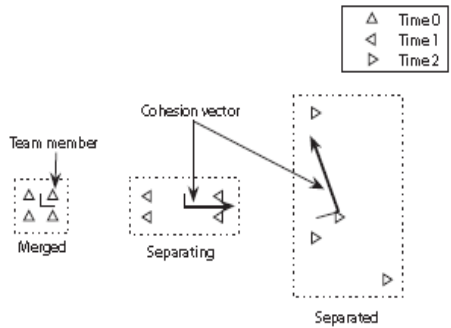

Figure 6: Discrete Team features [11]

## 2.6. Role Assignment / Learning (C5.0/ID3) and Mapping

In this stage each agent in the team is assigned to a role such that learning phase is done in a consistent and accurate way. Role assignment and mapping is done using C5.0, which is based on ID3 decision tree learning algorithm on GPU by Toby Sharp in [10]. The classification involves mapping the data structure that describes a decision forest to a 2D texture array. Navigation through the forest for each point of the input data is done in parallel. Regarding training, the responses of the training data are computed to a set of candidate features, then, these responses are scattered into a suitable histogram using a vertex shader. We integrated this framework getting it implemented on CUDA following [10].

## 2.7. TW activity Recognition / Learning (cuHMM)

Activity Recognition is the last phase in the framework that has the function of either learning or recognition according to the mode of the framework. One of the main learning models that have been used in action recognition with a great success is the Hidden Markov Model. Three kernel GPU implementation functions are provided: Forward, Viterbi and BalmWelch algorithms. HMM supported in this phase is not restricted only to teamwork actions but can be generally used in action recognition phase of any application. Under our framework, we integrated cuHMM by Chuan Liu in [4].

## 3. HIGH PERFORMANCE APPROACHES IN THE FRAMEWORK

Parallelism for applications under the presented framework can be achieved using the following approaches:

1. Component wise parallelism on GPU, in which the processing logic of each colored component in Figure 1 is distributed over GPU processing elements (in the scope of our experiments).

2. Sequential phases running on the framework can be pipelined (e.g. while detected objects of a frame are segmented in segmentation phase, the motion process is performed in the next frame).

3. Independent task parallelism, which can be exploited in object classification and action recognition phases. Object classification is a pure parallel task where multiple objects can be classified simultaneously. In action recognition HMMs are built for each action, which can run in parallel.

The methods above show how the parallelism is done under our framework. However availability of hardware increases the degrees of freedom for configuring the framework to gain more performance improvement. For instance, more GPU-tasks can be done by attaching more cards to the same PC, when applicable (High speed bus will be a critical issue here). In addition, the framework can be installed over multiple PCs connected by high speed LAN with each PC attached with one or more GPUs as known as GPUs-Grid.

## 4. Configuring Systems Running Under The Framework

There are two types of configurations required for full utilization of the GPU framework: one is software and the other is hardware.

### 4.1. Components' configuration (software)

In this stage, the developer makes selections of stages that will be utilized during the processing of his target system, and then adjusts the parameters of each component that fit into system requirements. The configuration parameters for each phase are listed as follows.

Moving Motion Detection By sliding window: length of sliding windows with 91 as the default length, and number of histogram pins [6]. Segmentation and objects' data extraction: the number of N partitions [7]. Object classification: feature vector length, and number of classes [5]. Mean Shift tracking: the number of clusters K as the tracking algorithm is based on K-means clustering [3]. Role Assignment and mapping (ID3 algorithm): mode (TreeLeaf, ForestLeaves and etc.), number of trees and number of classes [10].

Team Work activity Recognition (HMM): number of hidden states, feature vector size, blocks size and number of actions. The default value of the block size is 16 [4].

### 4.2. **Environment setup (hardware configuration)**

After components' configuration, the next is to install the hardware components in hand. The minimum requirement is to have one node with one GPU card on which all configured GPU components would run, the first approach in section 3.

For more complex situation where environment contains grid of nodes with each node containing at least a GPU card, all types of parallelism approaches in section 3 are applicable. Pipelining and Independent task parallelism are achieved either within same machine with multiple GPUs or across high speed LAN. The two methods are detailed as follows.

1. Multi-GPUs on a single machine: This is most recommended for tasks that require processing relatively large data to overcome communication cost over intranet. This will be more suitable for pipelining the stages of the configured systems (e.g. detection, tracking, segmentation, etc.).

2. GPU Grid distributed over different machines: This is recommended if a lot of learned data is stored over network and hence features would have relatively small size to broadcast over network then gain recognition data.

This would be suitable for classification and recognition that use large data for learning (e.g. object classification, action recognition phases). The key of successful configuration of the framework is how to organize available resources to fit processing requirements of the target system with optimal GPU utilization The optimum goal here is to configure the framework to minimize the processing time (denoted as T) required for recognizing actions given by Equation 1.

$$T = \sum PT_{cpu}(i) + \sum PT_{gpu}(j) + \sum CT_{gpu,cpu}(i,j) + \sum CT_{node,node}(i,j)$$

(Eq. 1)

Where $PT_{cpu}(i)$ is the processing time of task $i$ on a CPU. $PT_{gpu}(j)$ is the processing time of task $j$ on a GPU. $\sum CT_{gpu,cpu}(i,j)$ is the communication time for task k between a GPU $i$ and a CPU $j$. $CT_{node,node}(i,j)$ is the communication cost between two nodes $(i,j)$ belonging to the GPUs-Grid if applicable to perform the task.

## 5. EXPERIMENTAL RESULTS

This section provides speedup results for each of the GPU components comprising the framework. Besides, it provides overall speedup of the framework in terms of a selected application. In sub-sections 5.1 to section 5.6, we summarize the results of the components that we integrated into our framework as reported in their papers. Then in section 5.7, we present our experiment to evaluate the overall system performance in a teamwork activity recognition system, which achieves ~20X speedup on a very cheap GPU (NVidia 9500GT).

### 5.1. Moving Motion Detection by sliding window

GPU version using the adaptive mean as the background model with 91 frame sliding window can run at around 18 fps on 320x240 resolution videos. The mode approach can run at 10 fps with the same setting. The GPU version of the mean approach achieves around 12X speedup over its standard CPU counterpart. The GPU version of the mode approach achieves around 15x speedup [7].

### 5.2. Segmentation and objects' data extraction

This part is not recommended to use, due that its GPU version is a bit slower than CPU version. In case of image with 2048 x 2048 as dimensions on 9800GT, the speedup is 0.769231X, which means a slowdown [6].

However, it would result in speedup if GPUs with higher capabilities were used. So, it is subject to use depending on the available HW and complexity of input objects. Another reason is that the communication time compared to the processing time of the connected component algorithm is relatively high.

**Table1: Confusion matrix depicting teamwork activity recognition on UCF teamwork dataset**

|  | Actions | Traveling Column | Traveling line | Traveling box | Bounding Over-search | Wedge | Team Split | Team Merge | *Precision* |
|---|---|---|---|---|---|---|---|---|---|
| Predicted | Traveling Column | **29** | 0 | 0 | 0 | 0 | 0 | 0 | 100% |
| | Wedge | 0 | **17** | 0 | 0 | 0 | 0 | 0 | 100% |
| | Traveling box | 0 | 3 | **32** | 2 | 0 | 0 | 0 | 86.5% |
| | Bounding Over-search | 0 | 0 | 1 | **7** | 0 | 0 | 0 | 87.5% |
| | Traveling line | 0 | 0 | 0 | 0 | **26** | 0 | 0 | 100.0% |
| | Team Split | 0 | 0 | 0 | 0 | 0 | **13** | 2 | 86.7% |
| | Team Merge | 0 | 0 | 0 | 0 | 0 | 2 | **13** | 81.3% |
| | *Recall* | 100.00% | 86.7% | 97.0% | 77.8% | 100.00% | 86.7% | 86.7% | |

### 5.3. Object classification and Discretization (cuSVM)

Achieved speedup in cuSVM reached about 35x for training SVM, 84X for corresponding prediction. This speedup was recorded against MNIST data set (LeCun et al., 1998) on NVIDIA GTX 260 GPU [5].

### 5.4. Object tracking

The experiments show a speedup in blob tracking reaching 3.36x using the famous CAVIAR data set on GeForce 8800GTS GPU [3].

### 5.5. Role Assignment / Learning (C5.0/ID3) and Mapping

129x speedup in argmax mode using object recognition as an application on GPU NVidia GeForce GTX 280 (240 stream processors) [10].

### 5.6. TW activity Recognition / Learning (cuHMM)

880x speedup of forward algorithm GPU implementation while 180x is the speedup of Baumwelch algorithm GPU. This test was done on 512x512 (number of states x number of sequences) on NVIDIA G92 with 512MB RAM [4].

### 5.7. Evaluation of overall Framework Speedup

In order to test the applicability of the framework, we selected the system in [11] which is "role based activity recognition in observations of embodied agent actions". Linus Luotsinen et al in [11] used labeled data set, so motion detection, segmentation and tracking are filtered out in components' selection step of the configuration. The components' configuration of that system is presented as follows.

**Discretization**: in this component, the developer uses custom thresholds to transfer features from continuous to discrete.
**Role Assignment and mapping (ID3 algorithm):** Tree Leaf mode, single tree and 4 classes (4 different roles).
**TW activity Recognition (HMM):** number of hidden states is five, feature vector size is 6, and number of actions is six. .

This system was tested using UCF teamwork dataset on NVidia 9500GT graphics card (32 500MHZ processors) against 3.2GHZ dual core processor, and speedup of 19.97X was gained. The speedup was computed by comparing the time to generate the test results presented in Table one by GPU compared to CPU. Gained speedup increases dramatically as capabilities increases. We achieved almost the same results reported in [11].

The main gain here is the speedup with even a very cheap GPU card. In terms of accuracy we almost reproduced the results of [11] in Table 1.

## 6. CONCLUSION AND FUTURE WORK

We have presented a GPU framework for action recognition that can be configured to many action recognition systems;

furthermore key approaches to optimize utilization of the framework have been presented. Experiments show how the framework is configured in teamwork action recognition system with speedup of ~20x achieved on NVidia 9500GT graphics card.

Future work includes enrichment of each phase on the framework with more GPU implementations of relevant algorithms. Examples include GPU implementations of Particle filters for object tracking, Correleogram based segmentation algorithm to help discriminate overlapping objects, and Codebook based algorithm in object classification phase. In the future, we aim to extend this framework as a library for action recognition on GPU where state-of-the-art algorithms could be evaluated and used to speedup the performance of the intensive computations that is required for action recognition.